\documentclass[twocolumn,prc,showpacs,preprintnumbers,unsortedaddress,amsmath,amssymb,floatfix]{revtex4}
\usepackage{graphicx,color}
\usepackage{bm}
\usepackage{graphicx}
\usepackage{epsf}

\usepackage{comment}
\usepackage{color}
\usepackage{slashed}
\usepackage{ulem}

\newcommand{\delete}{\bgroup\markoverwith{\textcolor{red}{\rule[0.5ex]{2pt}{1pt}}}\ULon}

\newcommand{\be}{\begin{equation}}
\newcommand{\ee}{\end{equation}}
\newcommand{\bea}{\begin{eqnarray}}
\newcommand{\eea}{\end{eqnarray}}
\begin{document}

\title{Constraints on the Neutron Skin and the Symmetry Energy from the Anti-analog Giant Dipole Resonance in ${}^{208}$Pb}

\author{Li-Gang Cao$^{1,2,3,4}$, X. Roca-Maza$^{5,6}$,
G. Col$\grave{\text{o}}$$^{5,6,3}$, H. Sagawa$^{7,8,3}$}

\affiliation{${}^1$ School of Mathematics and Physics, North China Electric Power University, Beijing 102206, China}

\address{${}^2$ State Key Laboratory of Theoretical Physics, Institute of Theoretical Physics, Chinese Academy of Sciences, Beijing 100190, China}

\address{${}^3$ Kavli Institute for Theoretical Physics China, CAS, Beijing 100190, China}

\affiliation{${}^4$ Center of Theoretical Nuclear Physics, National
Laboratory of Heavy Ion Accelerator of Lanzhou, Lanzhou 730000,
China}

\affiliation{${}^5$Dipartimento di Fisica,
Universit$\grave{\text{a}}$ degli Studi di Milano, via Celoria 16,
20133 Milano, Italy}

\affiliation{${}^6$Istituto Nazionale di Fisica Nucleare (INFN),
Sez. di Milano, via Celoria 16, 20133 Milano, Italy}

\affiliation{${}^7$Center for Mathematics and Physics, University of
Aizu, Aizu-Wakamatsu, Fukushima 965-8580, Japan}

\affiliation{${}^8$RIKEN, Nishina Center, Wako, 351-0198, Japan}

\begin{abstract}
We investigate the impact of the neutron-skin thickness, $\Delta R_{np}$, on the 
energy difference between the anti-analog giant dipole resonance (AGDR), $E_{\rm AGDR}$, 
and the isobaric analog state (IAS), $E_{\rm IAS}$, in a heavy 
nucleus such as $^{208}$Pb. For guidance, we first develop a simple and analytic, yet physical, 
approach  based on the Droplet Model that linearly connects the energy difference 
$E_{\rm AGDR}-E_{\rm IAS}$ with $\Delta R_{np}$. To test this correlation on more fundamental 
grounds, we employ a family of systematically varied Skyrme energy density functionals 
where variations on the value of the symmetry energy at saturation density $J$ are explored. 
The calculations have been performed within the fully self-consistent Hartree-Fock 
(HF) plus charge-exchange random phase approximation (RPA) framework. We confirm 
the linear correlation within our microscopic apporach and, by comparing our 
results with available experimental data in $^{208}$Pb, we find that our 
analysis is consistent with $\Delta R_{pn}$ = 0.204 $\pm$ 0.009 fm, $J$ = 
31.4 $\pm$ 0.5 MeV and a slope parameter of the symmetry energy at saturation 
of $L$ = 76.4 $\pm$ 5.4 MeV --- the attached errors correspond to a lower-limit 
estimate of the systematic plus experimental uncertainties. These results are in agreement with those extracted from different experimental data albeit, $L$ and $\Delta R_{pn}$, are somewhat large when compared to previous estimations based on giant resonance studies.  
\end{abstract}


\maketitle

\section{Introduction}

Different experimental methods, either direct or indirect, have been proposed to extract the value of the neutron-skin thickness in finite nuclei, that is, the difference between neutron and proton root-mean-square radii,
\begin{equation}
\Delta R_{np} \equiv \langle r^2 \rangle_n^{1/2}
- \langle r^2 \rangle_p^{1/2}.
\end{equation}
The neutron skin thickness is an observable that has kept much attention from both experimental and theoretical viewpoints. This is because it is one of the most promising observables in nuclear structure to constrain the density dependence of the symmetry energy around the nuclear saturation density \cite{Tar14,Suz95,Kra91,Agr12,Roc11,Cen09,Vre03}. The symmetry energy plays an important role in understanding the mechanisms of different phenomena in nuclear physics and nuclear astrophysics \cite{Fur02,Hor01,Bro00,Die03,Tsa12,Dan02,Chen05,Chen10,Xu10,Yong06, Xiao09,Feng10,Gao12,Dong11,Li02,Riz08,New14,Lat14,Gan12,Kut94,Gai11}: it directly affects the properties of exotic nuclei, the dynamics of heavy-ion collisions, the structure of neutron stars, and the simulations of core-collapse supernova.

The Lead Radius Experiment (PREX) at the Jefferson Laboratory has provided the first 
model-independent evidence on the existence of a neutron-rich skin in $^{208}$Pb \cite{Abr12}. 
Relying on the fact that the weak charge of the neutron is much larger than the 
corresponding proton one, PREX used parity-violating electron scattering to probe 
the neutron distribution of $^{208}$Pb. To foster this field, more experiments 
have been already approved with both, the aim of improving the reached accuracy 
in ${}^{208}$Pb and to explore other mass regions. On the other side, neutron densities 
have been traditionally probed mostly by nucleon or $\alpha$ scattering. 
For example, by using proton elastic scattering on Sn and Pb isotopes \cite{Zen10}; 
or by measuring photons emitted during the decay of antiproton states \cite{Trz01,Klo07}. 
One can also obtain information on the neutron skin thickness from giant resonance properties, 
such as the excitation energy of the isovector giant dipole resonance (IVGDR), 
the total electric dipole polarizability ($\alpha_D$), the excitation energy of the 
isovector giant quadrupole resonance (IVGQR) or, yet with more warnings, from the energy and strength of the pygmy dipole resonance (PDR) in neutron-rich nuclei \cite{Adr05,Wie09,Tam11,Ros13,Kli07,Tri08,Cao08,Roc13a,Roc13,Car10,Pie11, Pie12,Zhang14}. Last but not least, the total strength of the charge-exchange spin-dipole resonances (SDR) can be related to the neutron skin in a very transparent way \cite{Vre03,Kra99,Yak05,Yak06,Sag07,Colo14,Loc14}. It is important to mention, however, that all hadronic probes require model assumptions to deal with the strong force introducing possible systematic uncertainties. 

Recently, the authors of Refs. \cite{Kra131,Kra132,Kra133} have proposed a new method to extract the neutron skin thickness based on the measurement of the excitation energy of the anti-analog giant dipole resonance (AGDR), that can be observed in the charge-exchange ($p,n$) reaction. The AGDR was first studied experimentally in Ref. \cite{Ste80}. Already in Ref. \cite{Krm83}, the authors had pointed out that the excitation energy of the AGDR is sensitive to the neutron-skin thickness. More recently, the energy difference between the AGDR and the isobaric analog state (IAS), $E_{\rm AGDR}-E_{\rm IAS}$, in $^{208}$Pb has been obtained by measuring the direct $\gamma$-decay between these states \cite{Yas13}. 

In this paper, we shall analyze the relationship of the neutron-skin thickness and the energy difference $E_{\rm AGDR}-E_{\rm IAS}$, by using a fully self-consistent Hartree-Fock (HF) plus charge-exchange random phase approximation (RPA) framework with a family of Skyrme energy density functionals. We try to understand also the qualitative features of such relationship through a simple, yet physical and transparent model. Our approach, as compared to Ref. \cite{Krm83}, incorporates specific effects of the Skyrme functionals such as the effective mass and the isovector enhancement factor (cf. Sec.~\ref{macsec}). Then, by comparing the theoretical and experimental results for $E_{\rm AGDR}-E_{\rm IAS}$, we extract the neutron-skin thickness in $^{208}$Pb. This allow us, in turn, to estimate the compatible values for the symmetry energy $J$ and its slope parameter $L$ (at nuclear matter saturation density). The extracted values of $J$ and $L$ are eventually compared to the results obtained by other analysis on different observables.

The outline of the paper is the following. In Sec.~\ref{micsec} the theoretical model is briefly presented: we focus, in particularly on the charge-exchange random phase approximation (RPA) based on the use of non-relativistic Skyrme energy density functional (EDFs). In Sec.~\ref{macsec}, we derive our analytic model to explain the relationship between the energy difference $E_{\rm AGDR}-E_{\rm IAS}$ of AGDR and IAS, and the neutron-skin thickness. A detailed quantitative analysis of such correlation is performed by employing a family of so-called SAMi-J Skyrme functionals, in Sec.~\ref{results}. In Sec.~\ref{conclusions} we summarize the results and draw our conclusions.

\section{Microscopic model: charge-exchange RPA}
\label{micsec}

The calculations are done within the framework of the Skyrme HF \cite{Bri72} plus charge-exchange RPA. We adopt the standard form of Skyrme interactions with the notations
of Ref.~\cite{Cha98}. Two nucleons characterized by the space, spin and isospin variables $\bm{r}_i$, $\bm{\sigma}_i$ and $\bm{\tau}_i$ interact through a zero-range, velocity-dependent and density-dependent force that reads
\begin{eqnarray}
V({\bm r}_1, {\bm r}_2) &=& t_0(1+x_0 P_{\sigma}) \delta({\bm r})
\nonumber \\ &+& \frac{1}{2} t_1 (1+x_1P_{\sigma})[{\bm
P}'^{2}\delta({\bm r})+
\delta({\bm r}){\bm P}^{2}] \nonumber \\
&+& t_2(1+x_2P_{\sigma}){\bm P}'\cdot\delta({\bm r}){\bm P}
\nonumber \\&+&
\frac{1}{6} t_3 (1+x_3P_{\sigma})\rho^{\alpha}({\bm R})\delta({\bm r})\nonumber\\
&+& iW_0 ({\sigma}_1+{\sigma}_2)\cdot[{\bm P}'\times\delta({\bf
r}){\bm P}]~, \label{eq2.1-1}
\end{eqnarray}
where $\bm{r}=\bm{r}_1-\bm{r}_2$, $\bm{R}=\frac{1}{2}(\bm{r}_1+\bm{r}_2)$, $\bm{P}=\frac{1}{2i}(\bm{\nabla}_1-\bm{\nabla}_2)$, $\bm{P}'$ is the hermitian conjugate of $\bm{P}$ (acting on the left), $P_{\sigma}=\frac{1}{2}(1+\bm{\sigma}_1\cdot\bm{\sigma}_2)$ is the spin-exchange operator, and $\rho=\rho_n+\rho_p$ is the total nucleon density. Within the standard formalism, the total binding energy of a nucleus can be expressed as the integral of the Skyrme density functional~\cite{Cha98}, which includes the kinetic-energy term $\mathcal{K}$, a zero-range term $\mathcal{H}_0$, the density-dependent term $\mathcal{H}_3$, an effective-mass term $\mathcal{H}_{eff}$, a momentum dependent term (that mimics finite-range effects) $\mathcal{H}_{fin}$, a spin-orbit term $\mathcal{H}_{so}$, a spin-gradient term $\mathcal{H}_{sg}$, and a Coulomb term $\mathcal{H}_{Coul}$.

Here, we will briefly summarize the formulas for the charge-exchange RPA calculations. The well-known RPA method~\cite{Ring80,Rowe} in matrix form is given by 
\bea\label{RPA} \left( \begin{array}{cc}
A & B \\
B^* & A^* \end{array}  \right) \left( \begin{array}{c}
X^\nu \\
Y^\nu  \end{array} \right) =E_\nu \left( \begin{array}{cc}
1 & 0\\
0 & -1 \end{array}  \right) \left( \begin{array}{c}
X^\nu \\
Y^\nu  \end{array} \right), 
\eea 
where $E_\nu$ is the energy of the $\nu$-th charge-exchange RPA state and  X$^\nu$, Y$^\nu$ are the corresponding forward and backward amplitudes, respectively. The matrix elements $A$ and $B$ are expressed as
\begin{eqnarray}
A_{mi,nj}=(\epsilon_m-\epsilon_n)\delta_{mn}\delta_{ij}+\langle{mj|V_{\rm res}|in}\rangle,
\end{eqnarray}
\begin{eqnarray}
B_{mi,nj}=\langle{mn|V_{\rm res}|ij}\rangle.
\end{eqnarray}
where the sub-indexes $i,j$ refer to occupied states, $m,n$ to unoccupied satates, $V_{\rm res}$ is the residual interaction and $\epsilon$ the single-particle states energy.

The particle-hole (p-h) matrix elements are obtained from the Skyrme energy density functional including all the terms (the Coulomb term $\mathcal{H}_{Coul}$ is not active in this case). The explicit forms of the matrices $A$ and $B$ are given in Ref.~\cite{colo13} in the case of a Skyrme force.

We will use the following operator for the AGDR excitation,
\begin{eqnarray}
\hat{O}_{\pm}=\sum_i
r_iY_{1m}(\hat{r}_i)t^{(i)}_{\pm},
\end{eqnarray} 
which corresponds to the $\Delta J=1$, $\Delta L=1$, $\Delta S=0$, $J^\pi=1^-$ resonance. 
We will also calculate the IAS. The IAS excitation operator reads
\begin{eqnarray}
\hat{O}_{IAS}=\sum_i t^{(i)}_{\pm},
\end{eqnarray}
and corresponds the $\Delta J=0$, $\Delta L=0$, $\Delta S=0$, $J^\pi=1^+$ states. 

\begin{figure}[hbt]
\includegraphics[width=0.45\textwidth]{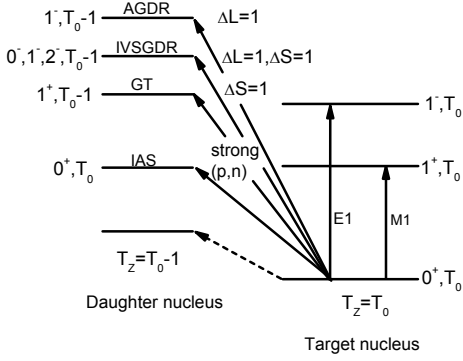}
\vglue -2.5cm \caption{Various states related to the target and daughter nucleus, including the ground state, the IVGDR and M1 excitations of the target nucleus (isospin = $T_0$), and the IAS (isospin = $T_0$), anti-analog states (isospin = $T_0-1$) in the daughter nucleus excited in a (p,n) reaction. } \label{Fig.1}
\end{figure}

In Fig. 1, we illustrate the relevant states of a target and its daughter nucleus, namely the ground state, the giant dipole resonance and the giant M1 state of a target nucleus. $T_0$ $\neq$ 0 labels the ground-state isospin of the target nucleus, that is, $(N-Z)/2$. The corresponding resonance states in the daughter nucleus reached by the ($p,n$) charge-exchange reaction are also displayed, namely the IAS (isospin = $T_0$), and the anti-analog (isospin = $T_0-1$) states: Gamow-Teller resonance (GTR), SDR, and AGDR. As shown in the figure, the AGDR corresponds to $\Delta J^\pi=1^-$, $\Delta L=1$, and $\Delta S=0$ excitation, and represents the anti-analog giant dipole resonance because it is the $T_0-1$ component of the charge-exchange of the isovector giant dipole resonance (IVGDR).

\section{Guideline from a simple analytic model}
\label{macsec}

\begin{table*}[t!]
\setlength{\tabcolsep}{.05in}
\renewcommand{\arraystretch}{1.3}
\caption{\label{tab2} Excitation energies and EWSRs ($m_0$)  of the AGDR and IVGDR for the family of SAMi-J interactions are displayed, together with the excitation energy of the IAS,  the enhancement factor $\gamma=\beta/(1+\alpha)$  and the prediction of the macroscopic model $E_{\rm AGDR}^{\rm mac}$ estimated from Eq.~(\ref{macmodel}). To evaluate $E_{\rm AGDR}^{\rm mac}$, we take $\varepsilon = 41 A^{-1/3} \sqrt{m/m^*}$ MeV,  $\Delta E_C =2\left(3/5\right)^{3/2} e^2 Z/\langle r^2\rangle^{1/2}$ and the energies of the current table. See the text for more details.}
\begin{tabular}{lccccccc}
\hline\hline
Force   & $E_{\rm AGDR}$ & $m_0^{\rm AGDR}$& $E_{\rm IVGDR}$
& $m_0^{\rm IVGDR}$& $E_{\rm IAS}$& $\gamma$ & $E_{\rm AGDR}^{\rm mac}$\\
 & [MeV]& [fm$^2$]& [MeV]& [fm$^2$]& [MeV]& &[MeV]\\
\hline
SAMi-J27 & 28.20 & 11153.99 & 13.99 & 4073.4 &  18.89 & 0.369 &25.51\\
SAMi-J28 & 27.94 & 11119.60 & 13.91 & 4071.0 &  18.74 & 0.366 &25.25\\
SAMi-J29 & 27.64 & 11079.18 & 13.74 & 4067.9 &  18.58 & 0.362 &24.87\\
SAMi-J30 & 27.37 & 11055.00 & 13.58 & 4065.4 &  18.42 & 0.360 &24.51\\
SAMi-J31 & 27.16 & 11047.33 & 13.42 & 4063.5 &  18.28 & 0.359 &24.18\\
SAMi-J32 & 26.94 & 11055.63 & 13.27 & 4062.6 &  18.16 & 0.361 &23.89\\
SAMi-J33 & 26.73 & 11078.24 & 13.13 & 4061.6 &  18.05 & 0.364 &23.62\\
SAMi-J34 & 26.54 & 11113.27 & 13.00 & 4061.3 &  17.96 & 0.368 &23.40\\
SAMi-J35 & 26.39 & 11158.82 & 12.88 & 4061.4 &  17.89 & 0.374 &23.21\\
\hline\hline
\end{tabular}
\end{table*}

In this Section we will develop a simple, yet physically sound, model for the excitation energies of the IVGDR, the IAS, and the AGDR. This effort will allow us to gain a deeper insight into the relevant ``macroscopic'' physics of our problem, namely the relationship between the neutron skin thickness and the energy difference of the AGDR and the IAS.

We start from the RPA dispersion relation for a separable interaction of the type $\kappa \hat{O}^\dagger\hat{O}$, that is,
\begin{equation}
\sum_{ph} \frac{\vert \langle p \vert \hat{O} \vert h \rangle
\vert^2}{\varepsilon_{ph}-E}+
\frac{\vert \langle p \vert \hat{O} \vert h \rangle
\vert^2}{\varepsilon_{ph}+E}
=-\frac{1}{\kappa},
\end{equation}
In this equation, $\varepsilon_{ph}$ is the unperturbed p-h excitation energy; if we assume that there is only one unperturbed configuration at energy $\varepsilon_{ph}=\varepsilon_0$ that exhausts the whole unperturbed strength $S_0$, then the equation becomes
\begin{eqnarray}
\frac{S_0}{\varepsilon_0-E} +\frac{S_0}{\varepsilon_0+E} =
-\frac{1}{\kappa},
\label{eq:RPA}
\end{eqnarray}
and, therefore,
\begin{equation}
E^2 = \varepsilon_0^2+2\kappa m_0  \label{eq:RPA2}
\end{equation}
where $m_0 =\varepsilon_0 S_0$ is the energy weighted sum rule (EWSR).

For each mode, there will be a different coupling constant; however, the isospin invariance dictates that the coupling constants of the AGDR and the non-charge exchange IVGDR to be the same. We start by considering the well-known IVGDR case: $\hat{O}=\sum_i r_i Y_{10}(\hat{r}_i) \tau_z^{(i)}$ --- note that for 
the macroscopic model we use the components of $\tau$ instead of $t$.  

We will use the Bohr-Mottelson quantal harmonic oscillator (QHO) model \cite{BM}; in this case, the coupling constant $\kappa$ for a dipole excitation is determined by the self-consistent condition between the vibrating potential and density:
\begin{eqnarray}
\kappa^{\tau=0}_{\lambda=1}=
-\frac{4\pi}{3 A}m\omega_0^2, \label{eq:is}\\
\kappa^{\tau=1}_{\lambda=1}=\frac{\pi V_1}{A\langle r^2\rangle}, \label{eq:iv}
\end{eqnarray}
for an isoscalar ($\tau=0$) or isovector ($\tau=1$) dipole ($\lambda=1$) excitations, respectively. $\hbar\omega_0$ is the major shell gap, $\approx$ 41 A$^{-1/3}$ MeV, $V_1$ is the strength of Lane potential $U=V_1 \frac{N-Z}{2A}$ and $\langle r^2 \rangle$ is the mean square radius. The classical EWSR of the isovector dipole mode is given by
\begin{equation}\label{eq:dipoleEWSR}
m_0(\tau=1, \lambda=1)= \frac{3}{4\pi}\frac{\hbar^2}{2m}A \ .
\end{equation}
It has to be noted that the energy $E^{\tau=0}_{\lambda=1}$ of the isoscalar dipole mode ({\it viz.}, the spurious center-of-mass mode) turns out to be at zero energy, as it should be. This can be checked from Eqs. (\ref{eq:RPA2}) and (\ref{eq:dipoleEWSR}). Moreover, the energy of the isovector giant dipole resonance, $E^{\tau=1}_{\lambda=1}\equiv E_{\rm IVGDR}$, becomes
\begin{equation}
E_{\rm IVGDR}^2 = \varepsilon_0^2 \left(1+
\frac{3\hbar^2V_1}{4 \langle r^2 \rangle m\varepsilon_0^2} \right),
\label{eq:RPA-IV}
\end{equation}
and it is well known that by using standard values for the unperturbed
energy ($\varepsilon_0=\hbar \omega_0=41/{\rm A}^{1/3}$ MeV) as well as for
the radius ($\langle r^2 \rangle = 3R^2_0/5=3(1.2)^2{\rm A}^{2/3}/5$ fm$^2$),
together with $V_1=130$ MeV, the excitation energy provided by Eq.
(\ref{eq:RPA-IV}) is $E^{\tau=1}_{\lambda=1} \approx 80/{A}^{1/3}$ MeV,
in  good agreement with the empirical systematics for the IVGDR
in the mass region ${\rm A} > 40$.

There are two important differences between the interaction assumed in the Bohr-Mottelson model and the Skyrme interaction. In the former case, the effective mass is taken to be $m^*/m$ = 1, while for Skyrme interactions this value depends on the chosen set, being in uniform matter as well as in the interior of nuclei close to the empirical value, $m^*/m\approx 0.7$. The effective mass changes the unperturbed energy to be $\varepsilon = \varepsilon_0 \sqrt{\frac{m}{m^*}}$. Moreover, in the case of momentum-dependent interactions such as the Skyrme forces (or other non-local forces), the classical EWSR should be multiplied by $1+\alpha$ where $\alpha$ is the so-called enhancement factor. For the dipole case, $\alpha$ is typically around $\approx$ 0.2
\cite{harakeh}.
Consequently, Eq. (\ref{eq:RPA2}) should turn into
\begin{equation}
E^2 =
\frac{m}{m^*}\varepsilon_0^2+2\kappa^\prime m_0 \left( 1+\alpha \right).
\label{eq:RPA3}
\end{equation}
It is a simple exercise to show that Eq. (\ref{eq:RPA3}) reproduces the experimental IVGDR systematics as well as Eq. (\ref{eq:RPA2}) if the coupling constant $\kappa^\prime$ is reduced with respect to $\kappa$, that is, $\kappa^\prime \approx 0.7\kappa$. By inspecting Eq. (\ref{eq:iv}) we can also conclude that this implies a quenched value for the strength of the Lane potential $V_1^\prime$ with $V_1^\prime \approx 0.7V_1$ (this value is of course indicative, in keeping with the rather crude approximations of this analytic model).

In the case of charge-exchange excitations of nuclei having a neutron excess, like ${}^{208}$Pb, the Tamm-Dancoff approximation (TDA) is known to provide results that are quite similar to those from the RPA. This is because the coupling between the $\tau_-$ excitations and $\tau_+$ excitations is small due to their quite large energy difference. We simplify thus the RPA dispersion relation to be the TDA one,
\begin{equation}
\sum_{ph} \frac{\vert \langle p \vert \hat{O} \vert h \rangle
\vert^2}{\varepsilon_{ph}-E} = -\frac{1}{\kappa}.
\end{equation}
As we have done already, we assume that there is only one unperturbed configuration at energy $\varepsilon_{ph}=\varepsilon_0$ exhausting the whole unperturbed strength $S_0$, so that
\begin{equation}
\frac{S_0}{\varepsilon_0-E} = -\frac{1}{\kappa}.
\label{eq:TDA}
\end{equation}
The solution of this simplified TDA equation is
\begin{eqnarray}
E &=& \varepsilon_0 + \kappa S_0 \nonumber \\
&=&\varepsilon_0 + \kappa \frac{m_0}{\varepsilon_0} \ . \label{eq:TDA2}
\end{eqnarray}

Let us consider the IAS first. It is well known that, to a first approximation, its excitation energy is associated with the Coulomb energy shift $\Delta E_C$ between the parent and the daugther nuclei. In our TDA model we can write the unperturbed $p-h$ energy as $\varepsilon_0 = -U+\Delta E_C$ where $U$ is, as above, the Lane potential. The non-energy weighted sum rule (NEWSR) obtained by using the operator $\sum_i \tau_-(i)$ is $2(N-Z)$. Therefore,
\begin{equation}
E_{\rm IAS} = -U + \Delta E_C + \kappa_{\lambda=0}^{\tau=1} 2(N-Z)
\label{eias}
\end{equation}
and if
\begin{equation}
\kappa= \frac{V_1}{4A}
\end{equation}
the IAS energy coincides with $\Delta E_C$.

Let us finally move to the subject of main interest for us, namely the AGDR.  Our goal is to have a transparent interpretation of the results obtained with the microscopic Skyrme model.

The AGDR has $\Delta L$ = 1, $\Delta S$ = 0, where the corresponding operator is
\begin{equation}
\sum_{i} r_i Y_{10} \tau_-(i).
\end{equation}
The NEWSR reads 
\begin{equation}
S_0(\tau_-,\lambda=1)-S_0(\tau_+,\lambda=1)=\frac{(N-Z)}{2\pi}\langle
r^2\rangle_{ne},
\end{equation}
where
\begin{equation}
\langle r^2 \rangle_{ne} \equiv \frac{N\langle r^2 \rangle_n
-Z \langle r^2 \rangle_p}{N-Z}.
\end{equation}
In this sum rule the $\tau_-$ contribution is largely dominant in nuclei with neutron excess like $^{208}$Pb; the same dominance holds for the energy-weighted sum rule, that can be written as 
\begin{equation}
m_0(\tau_{-},\lambda=1)-m_0(\tau_{+},\lambda=1)=
\frac{3}{2\pi}
\frac{\hbar^2 A}{2m}(1+\alpha+\beta).
\end{equation}
$\alpha$ is the same as in the IVGDR case that we have discussed above, whereas the definition of $\beta$ can be found, in the case of a Skyrme interaction, in Ref.~\cite{Aue81}.

Within the framework of our approximation, the AGDR unperturbed energy can be written as $\varepsilon-U+\Delta E_C$; consequently, its TDA energy from the simplified equation (\ref{eq:TDA2}) reads
\begin{eqnarray}\label{eq:agdr1}
E_{\rm AGDR} & = & \varepsilon - U + \Delta E_C 
+ \frac{V_1'}{2} \frac{\left( N\langle r^2 \rangle_n
-Z \langle r^2 \rangle_p \right)}{A\langle r^2\rangle},
\nonumber \\
& = & \varepsilon - U + \Delta E_C 
+ \frac{V_1'}{\langle r^2 \rangle} \frac{\frac{3}{2}
\frac{\hbar^2}{2m}(1+\alpha+\beta)}{\varepsilon - U + \Delta E_C}.
\nonumber \\
\end{eqnarray}
We are supposed to use the same coupling constant that has been already used in the case of the IVGDR; according to our previous discussion, this will be different from  the case of the Bohr-Mottelson model if used in conjunction either with an effective mass and/or with an enhancement factor as in the Skyrme case. For convenience, we shall define here $\bar{V}_1 \equiv V_1' (1+\alpha)$ --- note that $\bar{V}_1=V_1$ if $m^*/m=1$ and $\bar V_1 \approx V_1 $ even for realistic models with $m^*/m <1$. One can also notice that a simplification of Eq. (\ref{eq:agdr1}) comes from the fact that for a heavy nucleus such as ${}^{208}$Pb, replacing $\varepsilon - U + \Delta E_C$ with $\Delta E_C$ will produce an error of only a few \%. Specifically, if we assume $m*/m\approx 0.7$ and $V_1\approx 130$ MeV as previously done, $\varepsilon - U = 41 A^{-1/3} \sqrt{m/m^*} - V_1'(N-Z)/2A\approx 1.3$ MeV which correspond to about 7\% when compared to $\Delta E_C$. We use this simplification to write the
energy difference between the AGDR and the IAS:

\begin{eqnarray}
E_{\rm AGDR}-E_{\rm IAS} & = & 
\frac{\bar V_1}{2(1+\alpha)} \frac{\left( N\langle r^2 \rangle_n
-Z \langle r^2 \rangle_p \right)}{A\langle r^2\rangle},
\nonumber \\
& = & \frac{\bar V_1}{\langle r^2 \rangle} \frac{\frac{3}{2}
\frac{\hbar^2}{2m}(1+\alpha+\beta)}{\Delta E_C(1+\alpha)}.
\end{eqnarray}
It is convenient to define the quantity $\gamma\equiv\beta/(1+\alpha)$ since it is almost constant if we consider the interactions employed  in the current study (see Table \ref{tab2}). Finally, approximating the IAS energy as the Coulomb shift energy between  parent and daugther nuclei by $\Delta E_C=2\left(3/5\right)^{3/2}  e^2 Z/\langle r^2\rangle^{1/2}$, we may write
\begin{eqnarray}
E_{\rm AGDR} - E_{\rm IAS}&=& 
\frac{\bar V_1(1+\gamma)}{\Delta E_C}\frac{3}{2}\frac{\hbar^2c^2}{2mc^2\langle r^2\rangle}
\nonumber\\
&\approx& \frac{5}{8}\sqrt{\frac{5}{3}}\frac{\bar V_1 (1+\gamma)}{\alpha_{\rm H} Z}
\frac{\hbar c}{mc^2\langle r^2\rangle^{1/2}}.
\label{agdrias}
\end{eqnarray}
If we take $\alpha\approx 0.2$, $\gamma \approx 0.4$ and  $V_1 \approx 130$ MeV, we find $E_{\rm AGDR} - E_{\rm IAS}\approx 9$ MeV, which is in reasonable agreement with the result of our realistic calculations.

This schematic model gives us the opportunity to understand the sensitivity of $E_{\rm AGDR} - E_{\rm IAS}$ on the neutron skin thickness $\Delta R_{np}$. In fact, as it was done in Ref. \cite{Roc13} to which we confer the reader for details, we can relate the interaction strength of the potential $\bar V_1$ with the neutron skin thickness via the Droplet Model (DM),
\begin{equation}
\bar V_1 \approx 8\left[a_{\mathrm{sym}}(A) - \varepsilon_{F_\infty}/3\right].
\end{equation}
The DM also predicts that
\begin{eqnarray}
&&J-a_{\mathrm{sym}}(A)\approx
\frac{3J}{2\langle r^2\rangle^{1/2}}\frac{1}{I-I_C}\nonumber\\
&&\times \left(\Delta R_{np}-\Delta R_{np}^{\rm surf}+\frac{2}{7}I_C
\langle r^2\rangle^{1/2}\right),
\end{eqnarray}
where $I_C=e^2Z/20JR$ is a Coulomb correction to the total neutron excess $I=(N-Z)/A$, $a_{\rm sym}(A)$ is the symmetry energy parameter of the DM, and $\Delta R_{np}^{\rm surf}$ is a surface correction to the neutron skin thickness due to the different neutron and proton surface diffuseness. The latter quantity has been shown to be approximately constant in ${}^{208}$Pb ($\Delta R_{np}^{\rm surf}\approx 0.09\pm 0.01$ fm) when calculated by a large set of energy density functionals of different kind \cite{Cen10}. Since $I_C$ corresponds to a correction of about a 10\% to $I$ in heavy neutron-rich nuclei such as ${}^{208}$Pb, we will assume in what follows that $I-I_C\approx I$, and find 
\begin{eqnarray}
J-a_{\mathrm{sym}}(A)\approx\frac{3J}{2I}\times
\frac{\Delta R_{np}-\Delta R_{np}^{\rm surf}}{\langle r^2\rangle^{1/2}}
+\frac{3}{7}\frac{I_C}{I}J \ ,
\end{eqnarray}
and by combining this result with Eq.~(\ref{agdrias}) one finds that
\begin{eqnarray}
&&E_{\rm AGDR} - E_{\rm IAS}\approx 5\sqrt{\frac{5}{3}}
\frac{J}{I}\frac{1+\gamma}{\alpha_{\rm H} Z}
\frac{\hbar c}{m\langle r^2\rangle^{1/2}}\nonumber\\
&& \times\left[\left(1-\frac{\varepsilon_{F_\infty}}{3J}\right)I - 
\frac{3}{2}\left(\frac{\Delta R_{np}-\Delta R_{np}^{\rm surf}}
{\langle r^2\rangle^{1/2}}\right)-\frac{3}{7}I_C\right].\nonumber\\ 
\label{agdrias2}
\end{eqnarray}
For a given nucleus, Eq. (\ref{agdrias2}) predicts an explicit linear anti-correlation of $E_{\rm AGDR} - E_{\rm IAS}$ with $\Delta R_{np}$. We will show in the next Section that this correlation is actually displayed by the microscopic results.

We have also found very instructive to relate the different excitation energies within our macroscopic model, and check if the microscopic results follow such relationship. In doing that we have used the TDA expressions for the IVGDR, IAS and AGDR, and after some algebra, we arrive at
\begin{eqnarray}
E_{\rm AGDR} &=& \Delta E_C \left(1+\frac{\varepsilon - U}{\Delta E_C}\right)
\\&+& (E_{\rm IVGDR} - \varepsilon) 2 (1+\gamma)
\frac{\varepsilon}{\Delta E_C} \frac{1}{1+\frac{\varepsilon-U}{\Delta E_C}}
\ . \nonumber
\end{eqnarray}
As previously done, in a nucleus such as ${}^{208}$Pb, $(\varepsilon-U)/\Delta E_C$ can be neglected. Therefore, within a good approximation, we can write
\begin{equation}
E_{\rm AGDR} - E_{\rm IAS} \approx \frac{\varepsilon}{\Delta E_C}
\left(E_{\rm IVGDR} - \varepsilon\right) 2 (1+\gamma)
\end{equation}
or
\begin{equation}
E_{\rm AGDR} - E_{\rm IAS} \approx \frac{\varepsilon}{\Delta E_C}
\left(E_{\rm IVGDR} - \varepsilon\right) \frac{m_0^{\rm AGDR}}{m_0^{\rm IVGDR}}.
\label{macmodel}
\end{equation}
We define the energy of the AGDR extracted from  Eq. (\ref{macmodel}) as $E_{\rm AGDR}^{\rm mac}$, in Table I and hereafter. For the SAMi-J family and for a fixed nucleus, this formula suggests that the energy difference $E_{\rm AGDR} - E_{\rm IAS}$ should display the same trends as shown by $E_{\rm IVGDR}$. In fact, $\varepsilon$ depends only on the effective mass which is constant for the SAMi-J family ($\varepsilon \equiv 41 A^{-1/3} \sqrt{m/m^*}$ MeV). Moreover, $\Delta E_C$ is expected to not vary, and $1+\gamma$ is also approximately constant (cf. Table \ref{tab2}). The expression (\ref{macmodel}) reflects the idea that the physics encoded in the energy difference $E_{\rm AGDR} - E_{\rm IAS}$ reflects that of the IVGDR, as expected because of isospin invariance.

In Table \ref{tab2}, we present the predictions of the SAMi-J family for the different observables under study. The reader can verify that the latter equations of this Section reasonably reproduce the microscopic HF-RPA results: although there is an almost constant shift, the trend of $E_{\rm AGDR}$(RPA), however, is almost perfectly reproduced by the value  $E_{\rm AGDR}^{\rm mac}$ in Eq. (\ref{macmodel}). This finding gives us confidence in using the simple arguments in this Section to interpret the microscopic results. 

\section{Results and Discussions}
\label{results}

In this Section, we discuss the results obtained by employing the SAMi-J Skyrme energy density functionals to calculate the HF ground state and RPA excited states. The SAMi-J interactions are characterized by different values of the symmetry energy at saturation density: this value varies between 27 MeV and 35 MeV (in steps of 1 MeV), and the force parameters are fitted using properties of selected nuclei while keeping at the same time the constraints on few properties of nuclear matter (nuclear incompressibility $K_\infty$ = 245 MeV, and nucleon effective mass $m^*/m$ = 0.675). For details, the reader should consult Ref. \cite{Roc13}.

The ground state properties of $^{208}$Pb are calculated in coordinate space using box boundary conditions. The radius of the box is taken to be 20\,fm: the same box is used to calculate discrete states at positive energy that are associated with the continuum part of the spectrum. A cutoff energy of 60 MeV (in the single-particle energy) is adopted for the RPA calculations. With this energy cutoff, we have checked that the non-energy weighted sum rules for both AGDR and IAS are satisfied at the level of about 99.97\% for all Skyrme functionals used in the present study.

In Fig. 2, we show the response functions corresponding to the IAS and AGDR operators obtained for $^{208}$Pb by using the SAMi-J Skyrme functionals: the RPA results have been smeared out by using Lorentzian functions. As we can see,  the IAS the peak energy has small fluctuations as it varies between 17.5 and 18.6 MeV for the different SAMi-J parameter sets. As for the AGDR case, the peak energies vary between 26 MeV and 28.5 MeV by using the different SAMi-J parameter sets.  Experimentally, the mean AGDR energy has been extracted from the response function 
in the energy interval 5-15 MeV above the IAS energy.
To compare our results  with the experimental findings, we shall use the same energy range to calculate the mean energy from the AGDR response.

\begin{figure}[hbt]
\includegraphics[width=0.45\textwidth]{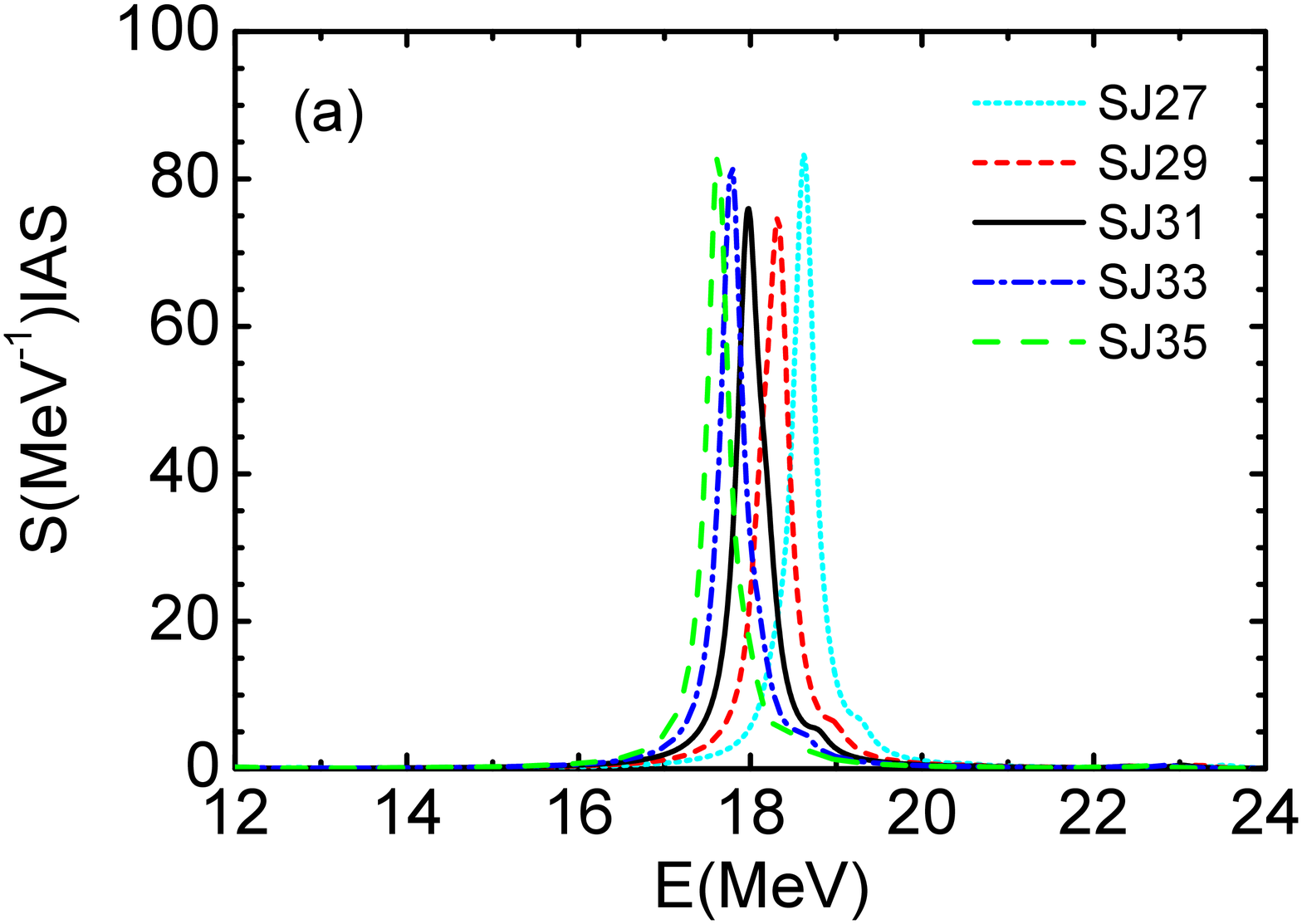}
\vglue -2.5cm
\includegraphics[width=0.45\textwidth]{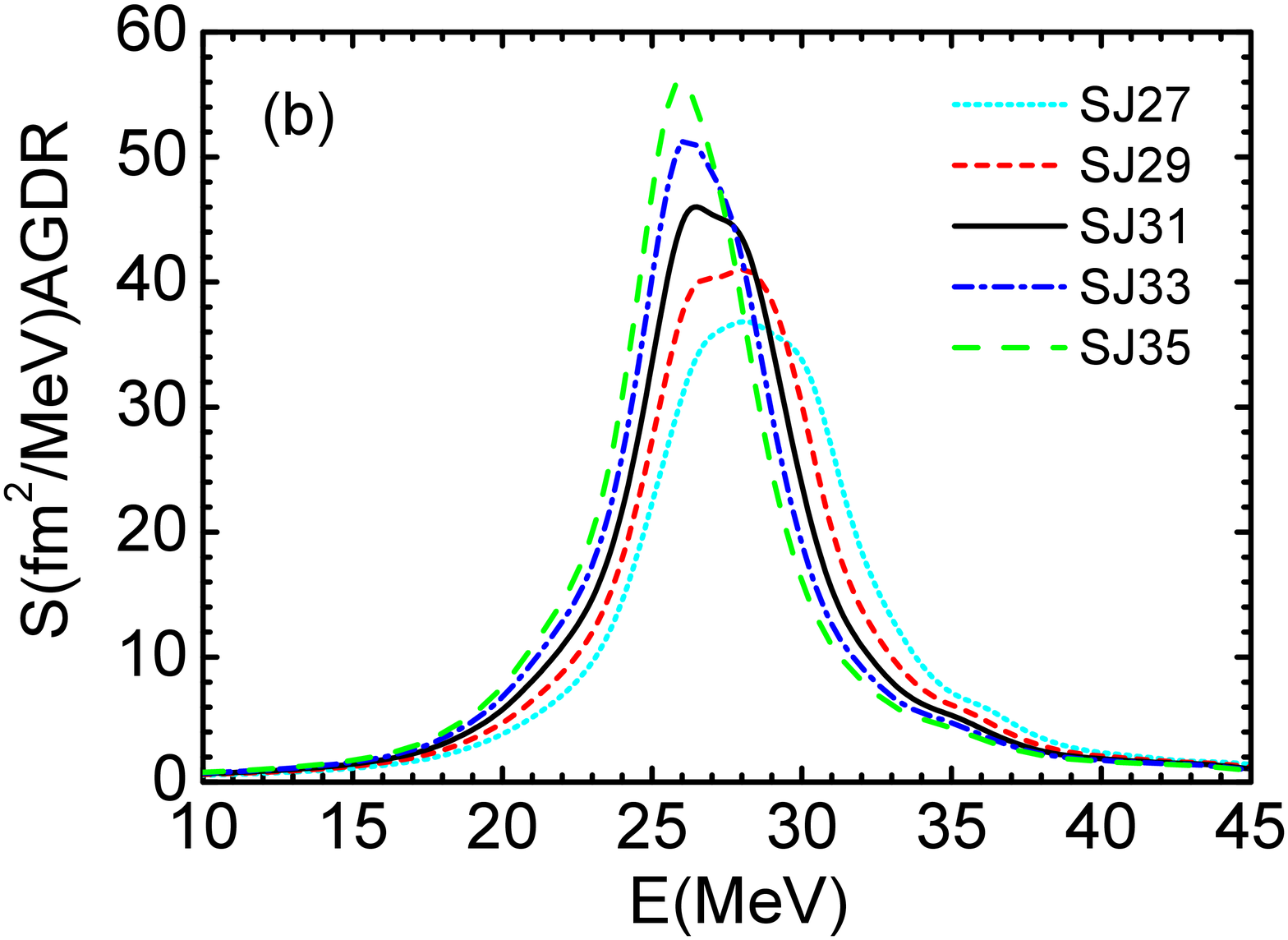}
\vglue -2.5cm \caption{(Color online) The (a) IAS and (b) AGDR response functions calculated 
by using the SAMi-J Skyrme energy density functionals. The discrete RPA peaks 
have been smeared out by using Lorentzian functions with (a) 300 keV and (b) 3 MeV width. 
} \label{Fig.2}
\end{figure}

In Fig. 3 we display the excitation energy of the AGDR and IAS as a function of the nucleon effective mass $m^*/m$, calculated with the SAMi-J and SAMi-m Skyrme functionals. We remind that the main difference between the SAMi-m and SAMi-J functionals is that in the former case the nucleon effective mass varies (in steps of 0.05) when fitting the parameters while $K_\infty$, $J$ and $L$ are kept constant (as above, we refer to \cite{Roc13} for details). The red squares in the panels correspond to results from SAMi-m. The results obtained from SAMi-J (black circles) are displayed in such a way that the bottom (top) point corresponds to the highest (lowest) value of $J$. The conclusion from these panels is that the excitation energy of the AGDR is sensitive to the symmetry energy at saturation density while, as expected, the variation of the AGDR excitation energy within the sets of the family SAMi-m is small. In the case of the IAS, the excitation energy is neither sensitive to the symmetry energy nor to the effective mass.

\begin{figure}[hbt]
\includegraphics[width=0.45\textwidth]{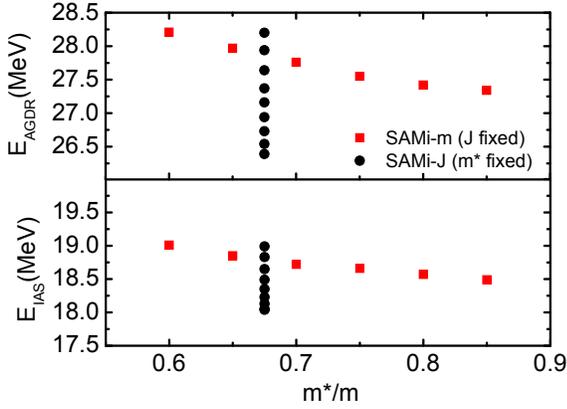}
\vglue -2.5cm \caption{(Color online) Excitation energies of the IAS and AGDR in $^{208}$Pb as a function of effective mass and symmetry energy at saturation density. The filled boxes are the results of the SAMi-m family, while the filled circles are those of the SAMi-J family. The top circle in the line of circles of each window corresponds  to the lowest $J$ value ($J$ = 27 MeV) and by going down one increases the $J$ value in steps of 1 MeV. The lowest circle then corrsponds to the maximum value, $J$ = 35 MeV.}
\label{Fig.3}
\end{figure}

\begin{figure}[hbt]
\includegraphics[width=0.45\textwidth]{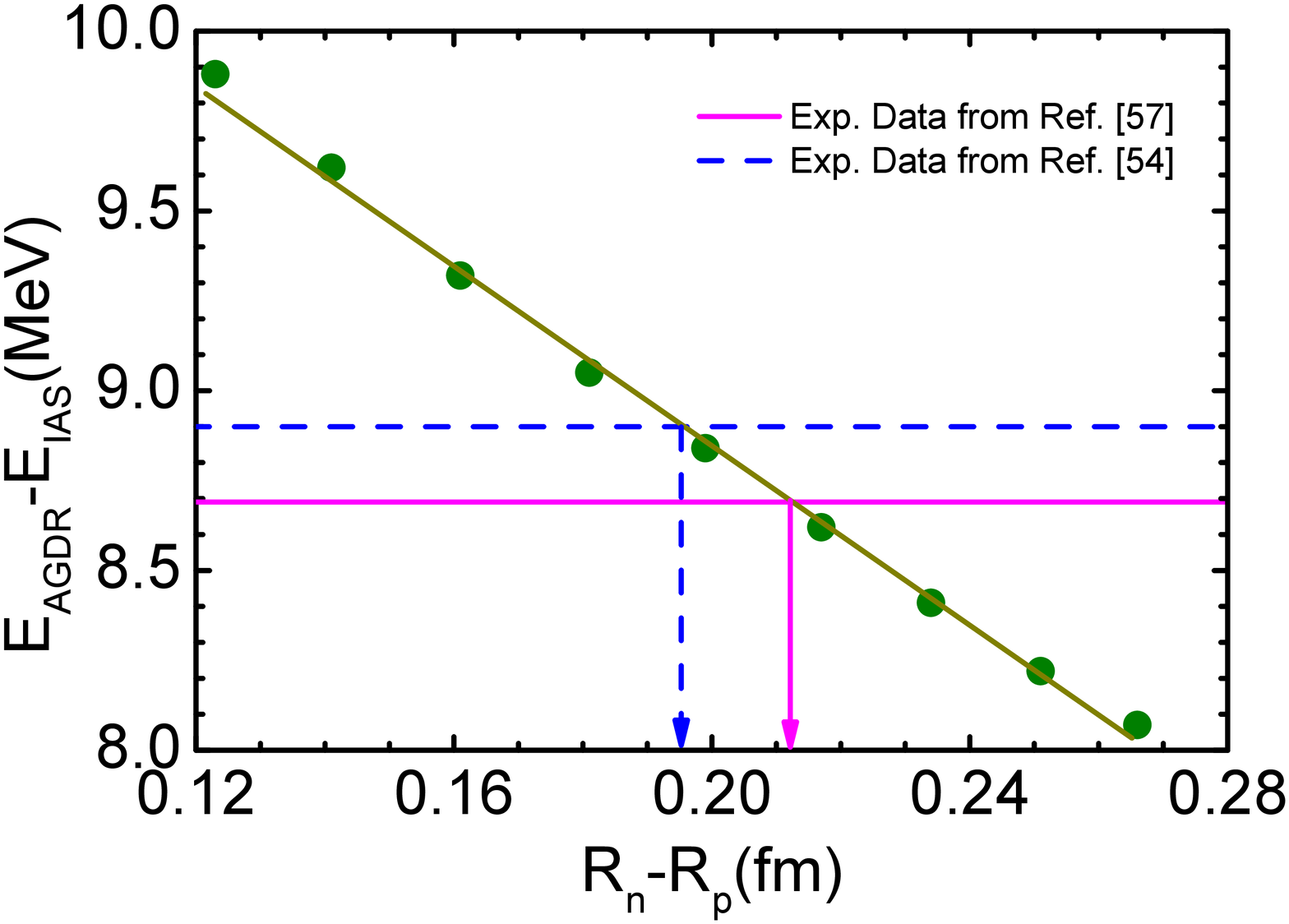}
\vglue -2.5cm \caption{(Color online) The energy difference $E_{\rm AGDR}-E_{\rm IAS}$ of AGDR and IAS as a function of neutron-skin thickness, obtained by using the SAMi-J family of Skyrme functionals consistently.  The calculated values are presented as solid circles. Two different experimental data \cite{Yas13,Kra133} are also shown as solid (magenta)  and dashed (blue) lines, respectively. The arrows indicate the neutron skin constrained by these experimental data.
} \label{Fig.4}
\end{figure}

\begin{table}[tp]
\setlength{\tabcolsep}{.05in}
\renewcommand{\arraystretch}{1.3}
\caption{\label{tab1} The values of the neutron-skin thickness of $^{208}$Pb obtained in the present work are compared to other values extracted by means of different experimental methods. }

\begin{tabular}{cccc}
\hline

 Method & Ref. & Date & $\Delta R_{pn}(fm)$  \\

antiproton absorption                      & \cite{Trz01} & 2001 & 0.180 $\pm$ 0.030 \\
($\alpha, \alpha'$) IVGDR                  & \cite{Kra04} & 2004 & 0.120 $\pm$ 0.070 \\
PDR                                        & \cite{Car10} & 2010 & 0.194 $\pm$ 0.024 \\
($\vec p,\vec{p'}$) & \cite{Tam11} & 2011 & 0.156 $\pm$ 0.025 \\
$\alpha_D$                                 & \cite{Pie12} & 2012 & 0.168 $\pm$ 0.022 \\
parity violation                           & \cite{Abr12} & 2012 & 0.330 $\pm$ 0.170 \\
($\gamma,\pi^0$)                           & \cite{Tar14} & 2014 & 0.150 $\pm$ 0.030 \\
AGDR                                       & present      & 2015 & 0.204 $\pm$ 0.009 \\
\hline
\end{tabular}
\end{table}

The calculated energy differences $E_{\rm AGDR}-E_{\rm IAS}$ between AGDR and IAS, 
obtained by employing the SAMi-J Skyrme functionals, are displayed as a function of the corresponding 
neutron-skin thickness in Fig. 4: in particular, the solid circles correspond to the sets SAMi-J27 
to SAMi-J35, from left to right. As we mentioned above, for the excitation energy of the AGDR 
we take the centroid of the theoretical strength distribution, calculated in 
the energy interval from 5 to 15 MeV above the IAS energy. 
The results show that the energy differences $E_{\rm AGDR}-E_{\rm IAS}$ between AGDR and IAS decrease with increasing values of the neutron-skin thickness, and a strong linear correlation exists; this is quite well justified by the model that has been developed in Sec.~\ref{macsec}.

In Fig. 4 we also super-impose two different experimental data. In Ref. \cite{Yas13} (that will be denoted as Exp1 hereafter), the AGDR has been separated from other excitations by means of the multipole decomposition analysis of the $^{208}$Pb($\vec{p},\vec{n}$) reaction at a bombarding energy $T_p=296$ MeV: the polarization transfer observables have been, in this case, quite instrumental to separate the non-spin flip AGDR from the spin-flip SDR in the multipole decomposition analysis. The energy difference between the AGDR and the IAS was determined to be $E_{\rm AGDR}-E_{\rm IAS}$ = 8.69 $\pm$ 0.36 MeV, where the uncertainty is claimed to include both statistical and systematic contributions. We show this datum by a solid (magenta) line in Fig. 4. The other experimental measurement has been reported in Ref. \cite{Kra133} (Exp2): in this case, the $^{208}$Pb($p,n\gamma p$) $^{207}$Pb reaction at a beam energy of 30 MeV has been used to excite the AGDR and to measure its $\gamma$-decay to the isobaric analog state, in coincidence with proton decay of the IAS. The energy difference $E_{\rm AGDR}-E_{\rm IAS}$ between the AGDR and the IAS was determined to be $E_{\rm AGDR}-E_{\rm IAS}$ = 8.90 $\pm$ 0.09 MeV. We show this result by means of a dashed (blue) line in Fig. 4. Given the error bars, the two work provide consistent results.

\begin{figure}[hbt]
\includegraphics[width=0.45\textwidth]{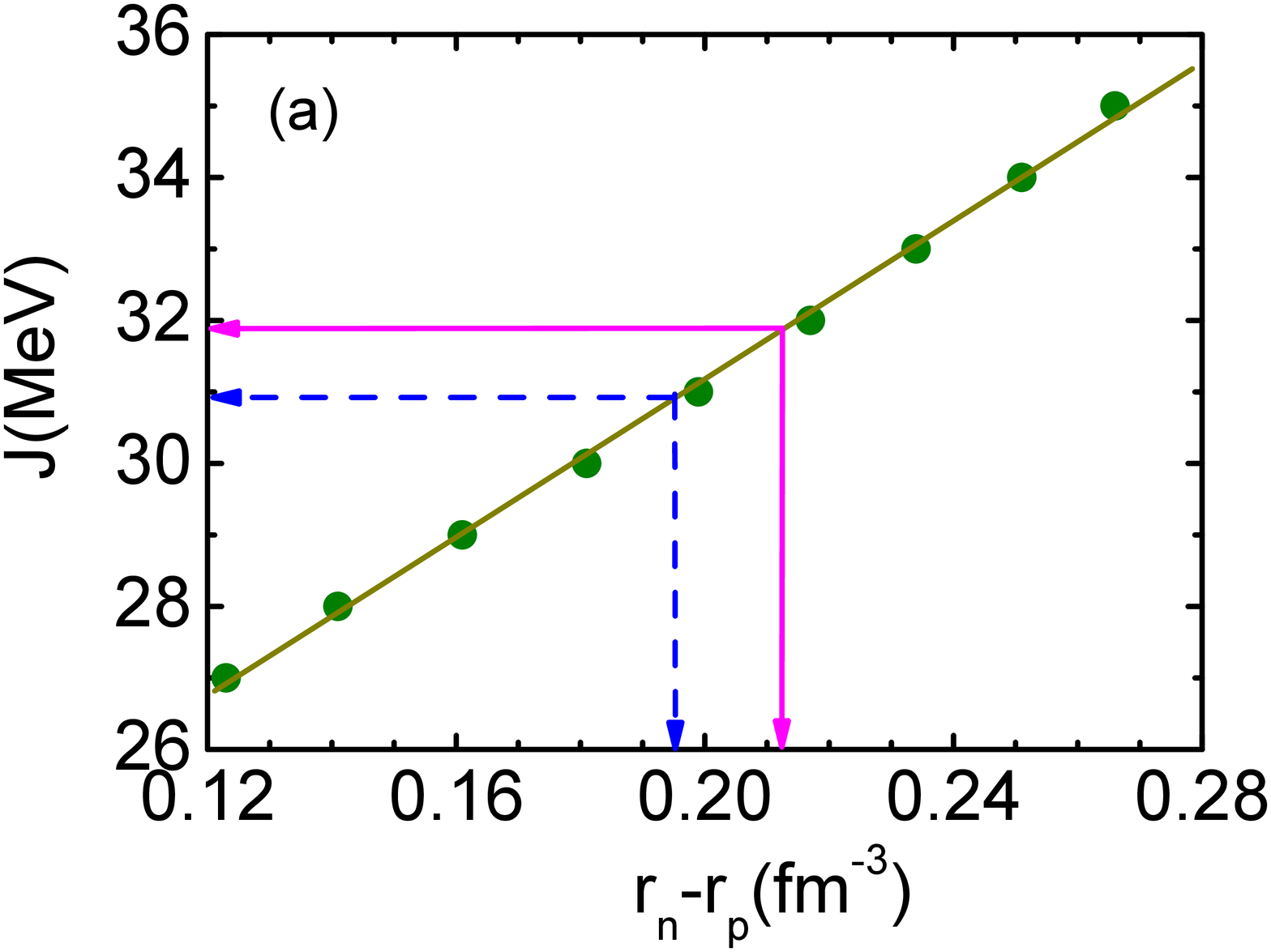}
\vglue -2.5cm
\includegraphics[width=0.45\textwidth]{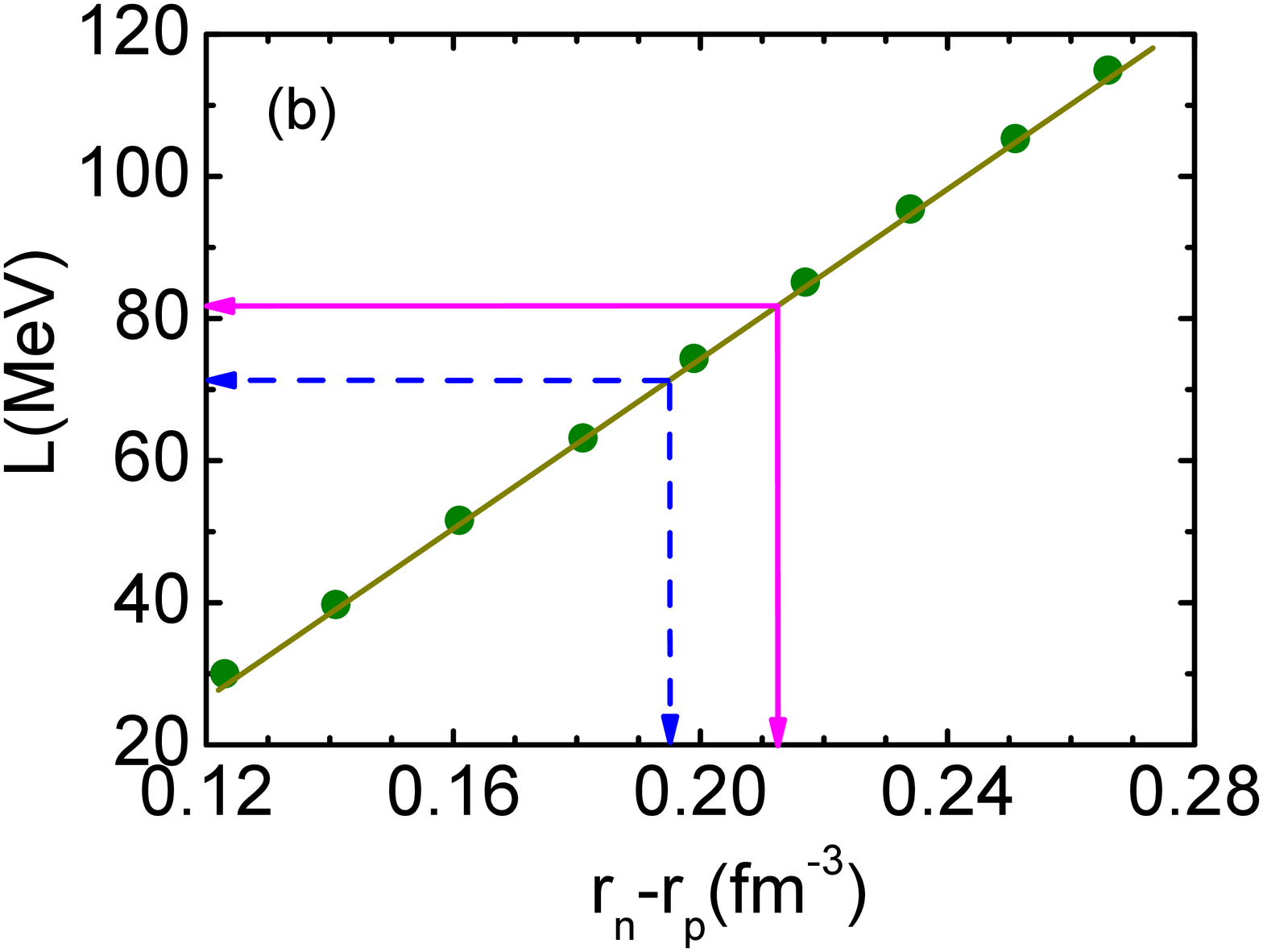}
\vglue -2.5cm
 \caption{(Color online) The (a) upper and (b) lower panels
show the correlations between the neutron-skin thickness and
either the symmetry energy $J$ at saturation density or the
corresponding slope parameter $L$, respectively. The constraints
provided by the experimental data already shown in Fig. 4.
} \label{Fig.5}
\end{figure}

\begin{figure}[hbt]
\includegraphics[width=0.45\textwidth]{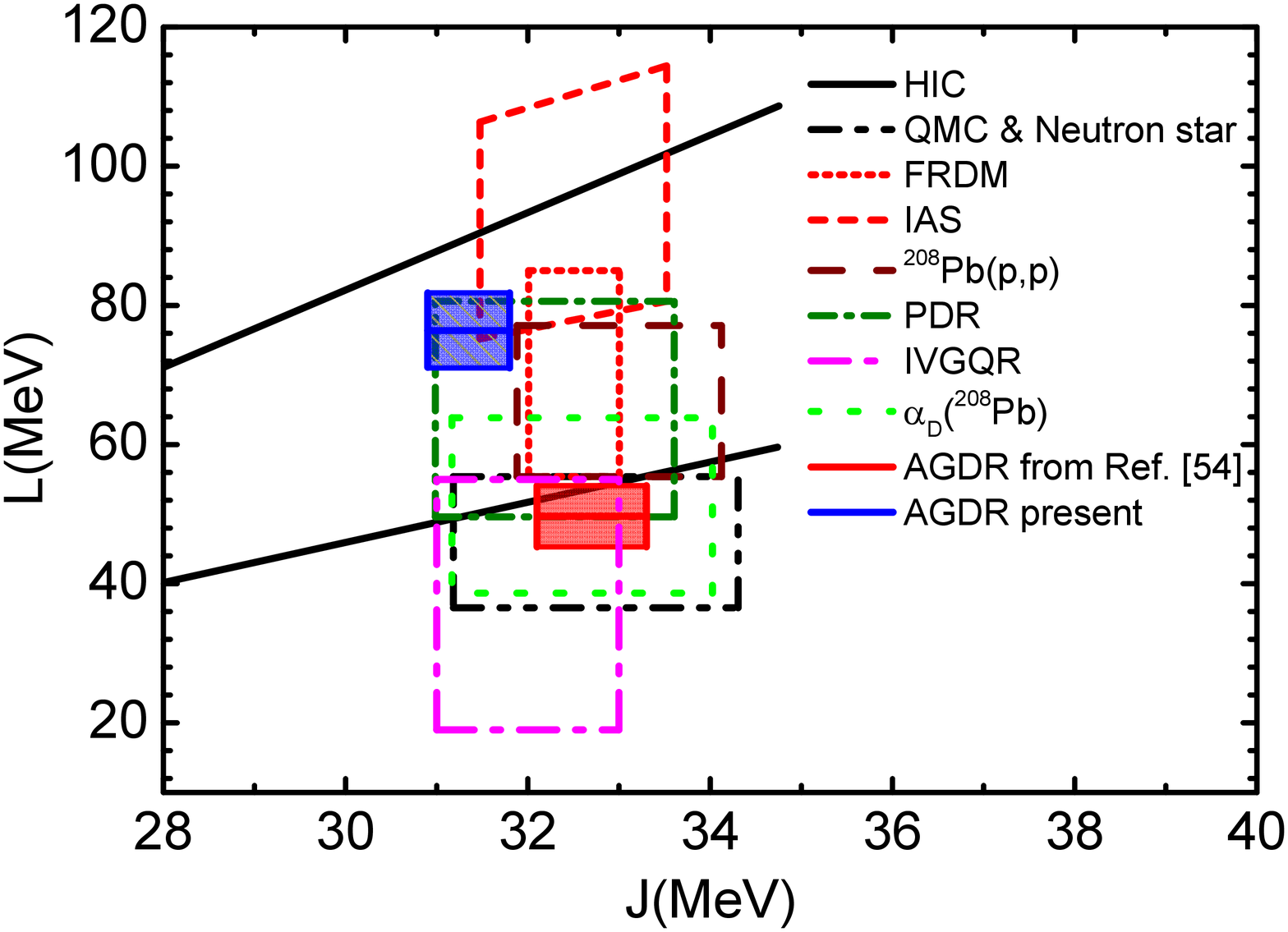}
\vglue -2.5cm \caption{(Color online) The values of the slope parameter
$L$ and symmetry energy $J$ at saturation density extracted in the
current work are compared with the values extracted from other
experimental data with several different methods. } \label{Fig.6}
\end{figure}

By comparing the experimental data for the energy difference of AGDR and IAS with our correlation line, we find that the value of the neutron-skin thickness of $^{208}$Pb is $\Delta R_{pn}$ = 0.212 $\pm$ 0.028 fm using Exp1, and $\Delta R_{pn}$ = 0.195 $\pm$ 0.007 fm using Exp2, respectively: these values are indicated by arrows in Fig. 4. The weighted average of these two results, that is, 
0.204 $\pm$ 0.09 fm for $\Delta R_{pn}$, is reported in Table I. We also compare this result 
with previous results deduced from different experimental methods. Although on the high side, our result is compatible with several other estimates. In Ref. \cite{Yas13,Kra133}, the authors also compared their experimental data with the values of $E_{\rm AGDR}-E_{\rm IAS}$ of AGDR and IAS obtained by using a fully self-consistent proton-neutron relativistic RPA with a family of density-dependent meson-exchange interactions (DD-ME) \cite{Vre031}.  Including the uncertainty both from experimental and theoretical sides, finally they found the value of the neutron-skin thickness of $^{208}$Pb to be $\Delta R_{pn}$ = 0.216 $\pm$ 0.046 fm $\pm$ 0.015 fm for Exp1, and $\Delta R_{pn}$ = 0.190 $\pm$ 0.028 fm for Exp2, respectively: these agree well with our results obtained using non-relativistic Skyrme energy density functionals.

We have also extracted the symmetry energy $J$ and its slope parameter $L$ at saturation density 
by using the neutron-skin thickness presently obtained. The results are shown in Fig. 5. 
The value for symmetry energy $J$ is extracted to be $J$ = 31.8 $\pm$ 1.6 MeV ($J$ = 30.9 $\pm$ 0.5 MeV) 
from Exp1 (Exp2) at the saturation density, and the value for the slope parameter $L$ 
of symmetry energy at saturation density is $L$ = 81.8 $\pm$ 17 MeV ($L$ = 71 $\pm$ 4 MeV) 
for Exp1 (Exp2). The weigthed average of these results is 
$J$ = 31.4 $\pm$ 0.5 MeV and $L$ = 76.4 $\pm$ 5.4 MeV. Of course, by making the weighted average one reduces the error bars, and this may hide even further systematic differences between the experiments and/or model dependences. As we mentioned in the abstract, the reported errors correspond to a lower-limit estimate of the systematic plus experimental uncertainties. 

In Fig. 6, the extracted values of $J$ and $L$ by the present analysis are shown together with those 
obtained with other  methods. These include: Quantum Monte Carlo (QMC) simulations 
of neutron stars \cite{Ste12}, analysis of the nuclear binding energies (by FRDM) \cite{Pol12}, 
energies of isobaric analog states (IAS) \cite{Dan09}, proton elastic scattering ($^{208}$Pb (p,p)) 
\cite{Zen10}, pygmy dipole resonances (PDR) \cite{Car10}, total dipole polarizability \cite{Roc13a}, 
and excitation energy of the isovector giant quadrupole resonance \cite{Roc13}. We should note that the presently extracted value of $J$ is consistent  with the other values in Fig. 6, with small variations. On the other hand, the present value of $L$,  although similar to those from the IAS analysis and binding energies from FRDM, is somewhat larger than the average value of all other deductions.

In Ref. \cite{Kra133}, the values of $L$ and $J$ were extracted by using the same experimental energy difference of AGDR and IAS that we have used (Exp2). They have obtained $J$ = 32.7 $\pm$ 0.6 MeV and $L$ = 49.7 $\pm$4.4 MeV, as shown in Fig. 6 with red shaded area. While our result for $J$ is consistent with the one obtained in Ref. \cite{Kra133}, the present (central) value of $L$ is about 40\% larger than that of Ref. \cite{Kra133}. This may be due to the diffenent energy density functionals used in the present analysis and in Ref. \cite{Kra133}, where the RMF Lagrangians of DD-ME type were adopted. The inputs for fitting the DD-ME and SAMi-J functionals are not exactly the same, and a different ansatz for the density dependence is assumed. This shows up in, e.g., different values for the nuclear incompressibility that turns out to be $K_\infty$ = 270 MeV in the relativistic case and $K_\infty$=245 MeV for the non-relativistic SAMi family. However, a clear explanation of this difference is a point that remains for future study.

We should also notice that in the previous studies devoted to the extraction of $L$ and $J$ 
from giant resonances, we have found values of $L$ like 64.8 $\pm$ 15.7 from PDR 
and 37 $\pm$ 18 from IVGQR as reviewed in \cite{Colo14}. 
These values are smaller than the present value: the present value is consistent with one
of the previous estimates but not with both of them. On the other hand, $J$ is consistent with other extractions from giant resonance data.

\section{SUMMARY AND PERSPECTIVE}
\label{conclusions}

In this work, we have studied the correlation of the neutron-skin thickness and the energy 
difference $E_{\rm AGDR}-E_{\rm IAS}$ of AGDR and IAS in $^{208}$Pb, by using a family of effective Skyrme energy density functionals, named SAMi-$J$ (SAMi-$m$), that are characterized by different values of symmetry energy $J$ (effective mass $m^*$). The calculations have been done within a fully self-consistent Skyrme HF plus charge-exchange RPA framework. We find a strong linear correlation of the energy difference $E_{\rm AGDR}-E_{\rm IAS}$ with the neutron-skin thickness $\Delta R_{pn}$ in $^{208}$Pb. An analytic model has been developed to explain the dependence of the excitation energy of AGDR on the neutron-skin thickness, in which it becomes apparent that such excitation energy decreases when the neutron-skin thickness increases. We also confirmed that the symmetry energy $J$ and the slope parameter $L$ have linear correlations with the the neutron-skin thickness within the employed Skyrme SAMi-J models.

Accordingly, 
we have extracted the neutron-skin thickness in $^{208}$Pb as 
$\Delta R_{pn}$=0.204$\pm$0.009 fm by comparing with the corresponding experimental 
energy difference of AGDR and IAS. Finally, we have also constrained the symmetry energy 
($J$ = 31.4 $\pm$ 0.5 MeV) and its slope parameter ($L$ = 76.4 $\pm$ 5.4 MeV) at saturation density by using the value of the neutron skin. Good agreement is obtained in comparing with our new results for the neutron-skin thickness and the symmetry energy $J$ with the values extracted with many different experimental methods. On the other hand, the presently extracted $L$ value is somewhat larger than the previously obtained values. The reported errors in our theoretical analysis correspond to a lower-limit estimate of the systematic plus experimental uncertainties. 

The use of the ($p,n$) reaction to study the AGDR can be extended to unstable nuclei due to the progress made in the development of new experimental techniques involving radioactive beams in inverse kinematics \cite{Sas11,Sas12}. Further experimental efforts on the AGDR in other mass regions and/or in long isotopic chains are desirable to increase the predictive power of current energy density functionals and to reduce the model dependence that one deals with when extracting nuclear matter properties. This may eventually allow us to better constrain the equation of state of asymmetric nuclear matter, a landmark for nuclear physics and nuclear astrophysics.

\section*{ACKNOWLEDGEMENTS}
This work is supported by the National Natural Science Foundation of China under Grant Nos 11175216 and 11435014, and the Fundamental Research Funds for the Central Universities (JB2014241).  This work is also supported by the Japanese Ministry of Education, Culture, Sports, Science and Technology by a Grant-in-Aid for Scientific Research under the program number (C) 22540262.

\end{document}